\documentclass[a4paper,11pt]{article}
\usepackage{pos}
\usepackage[nolist,nohyperlinks]{acronym}
\usepackage{caption}
\usepackage{subcaption}
\usepackage{float}
\usepackage{tikz}

\usetikzlibrary{decorations.markings,decorations.pathmorphing,positioning,decorations.pathreplacing,shapes,calc}
\tikzset{
    photon/.style={decorate, decoration={snake,amplitude=1pt,segment length=6pt}},
    zigzag/.style={decorate, decoration=zigzag},
}

\makeatletter
    \newacro{EW}{electroweak}
    \newacro{DET}{disperon effective theory}
    \newacro{EFT}{effective field theory}
    \newacroplural{EFT}[EFTs]{effective field theories}
    \newacro{sQED}{scalar QED}
    \newacro{FsQED}{form-factor scalar QED}
    \newacro{RGE}{renormalisation group evolution}
    \newacro{LL}{leading logarithmic}
    \newacro{NLL}{next-to-leading logarithmic order\def\AC@acronyms@LL{@<>@<>@}}
    \newacro{SM}{Standard Model}
    \newacro{HVP}{hadronic vacuum polarisation}
    \newacro{LO}{leading order}
    \newacro{NLO}{next-to-leading order\def\AC@acronyms@NNLO{@<>@<>@}\def\AC@acronyms@LO{@<>@<>@}\def\AC@acronyms@NNNLO{@<>@<>@}}
    \newacro{NNLO}{next-to-next-to-leading order\def\AC@acronyms@NNNLO{@<>@<>@}\def\AC@acronyms@NLO{@<>@<>@}\def\AC@acronyms@LO{@<>@<>@}}
    \newacro{NNNLO}[N$^3$LO]{next-to-next-to-next-to-leading order\def\AC@acronyms@NLO{@<>@<>@}\def\AC@acronyms@LO{@<>@<>@}\def\AC@acronyms@NNLO{@<>@<>@}}
    \newacro{VFF}{vector form factor}
    \newacro{LEFT}{low-energy effective field theory}
    \newacro{IR}{infrared}
    \newacro{ISC}{initial-state corrections}
    \newacro{FSC}{final-state corrections}
    \newacro{FKSl}[FKS$^\ell$]{}
    \newacro{NTS}{next-to-soft}
\makeatother

\renewcommand{\Im}{\mathrm{Im}}

\def\mcmule{{{\sc McMule}}}
\def\thavg{\theta_{\rm avg}}
\newcommand{\OpenLoops}{{\rmfamily\scshape OpenLoops}}

\title{Beyond QED: \\ Electroweak and hadronic extensions of McMule}
\ShortTitle{Beyond QED: Electroweak and hadronic extensions of McMule}

\author*[a,b]{Sophie Kollatzsch}

\affiliation[a]{PSI Center for Neutron and Muon Sciences, \\ Forschungsstrasse 111, 5232 Villigen PSI, Switzerland}

\affiliation[b]{Physik-Institut, Universit\"at Z\"urich, \\ Winterthurerstrasse 190, 8057 Z\"urich, Switzerland}

\emailAdd{sophie.kollatzsch@psi.ch}

\abstract{
\mcmule{} is a Monte Carlo framework developed to advance the low-energy precision frontier by providing QED corrections to leptonic scattering and decay processes, currently up to \acl{NNLO}.
Recent developments have extended its capabilities in two important directions: the systematic inclusion of \acl{EW} effects at low energies within the \acl{LEFT}~\cite{Kollatzsch:2025pnp}, and the incorporation of pion form factors and the non-perturbative hadronic vacuum polarisation in loop amplitudes through a combination of \OpenLoops{} and effective field theory techniques, referred to as disperon QED~\cite{Fang:2025mhn}.
I will provide an overview of \mcmule{} and discuss these recent extensions and their applications. 
In particular, I will illustrate the impact of the model used for non-perturbative $\gamma$–$Z$ mixing effects in the context of the MOLLER experiment and highlight the subtlety involved in consistently aligning \OpenLoops{} with its effective field theory expansion in disperon QED.
}

\FullConference{17th International Symposium on Radiative Corrections: Applications of Quantum Field Theory to Phenomenology (RADCOR2025)\\
5-10 October 2025\\
Puri, India\\}

\begin{document}
\maketitle
\section{Introduction}
Low-energy scattering processes provide an exceptionally sensitive laboratory for exploring the \ac{SM}, enabling the extraction of its parameters and the investigation of fundamental questions in particle physics.
The success of this endeavour hinges on the availability of precise Monte Carlo tools.
While several such codes exist (see, for example, the community report~\cite{Aliberti:2024fpq}), in the following we focus exclusively on \mcmule{}~\cite{Banerjee:2020rww,mcmule:website}.

Motivated by experiments such as MUonE~\cite{CarloniCalame:2015obs,MUonE:2016hru}, demanding predictions at the \ac{NNLO} level and beyond, substantial progress has been made in QED calculations.
In parallel, increased attention has been devoted to contributions beyond pure QED, in particular \ac{EW} and hadronic effects~\cite{Aliberti:2024fpq,Afanasev:2023gev}.
While the \ac{IR} structure remains unchanged compared to pure QED, \ac{EW} and hadronic effects introduce additional conceptual and calculational challenges.
Some of them can be addressed using \acp{EFT}, which provide a systematic and efficient description of physical systems with hierarchical scales.

In these proceedings, we briefly review \mcmule{} and its core principles that allow for QED predictions at \ac{NNLO} and beyond in Section~\ref{sec:mcmule}.
In the following Section~\ref{sec:recent}, we discuss its extensions to
hadronic effects using a purpose-built \ac{EFT} in Section~\ref{sec:hadrons}
and to
\ac{EW} physics via \ac{LEFT} in Section~\ref{sec:EW}, before presenting plans for future work in Section~\ref{sec:conclusions}.

\section{Overview of McMule}
\label{sec:mcmule}
Taking as input the required matrix elements, the Monte Carlo framework \mcmule{} provides fully-differentiable calculations to leptonic scattering and decay processes.
\mcmule{} is built around three central aspects:
\begin{itemize}
    \setlength\itemsep{0em}
    \item handling the \ac{IR} structure of QED with physical masses. Due to the simple exponentiation of \ac{IR} singularities, the all-order FKS$^\ell$ subtraction scheme~\cite{Engel:2019nfw} can be employed.
    \item computing two-loop matrix elements with non-vanishing masses.
    If the mass is the smallest scale of the process, the leading mass effects can be recovered using massification~\cite{Penin:2005eh,Becher:2007cu,Engel:2018fsb,Bonciani:2021okt}.
    \item maintaining numerical stability. In problematic regions of the phase space where the photon becomes too soft, the full matrix element is replaced by a numerically adequate expression up to $\mathcal{O}(E_\gamma^0)$~\cite{Banerjee:2021mty}.
\end{itemize}
\mcmule{}, heavily relying on \OpenLoops{}~\cite{Buccioni:2017yxi,Buccioni:2019sur}, is able to provide QED predictions at \ac{NNLO} for several scattering processes such as $\mu e \to \mu e$~\cite{Broggio:2022htr}, $ee\to \mu\mu$~\cite{Kollatzsch:2022bqa,Aliberti:2024fpq}, $ee\to ee$~\cite{Banerjee:2021qvi,Banerjee:2021mty} and $\ell p \to \ell p$~\cite{Engel:2023arz}.

\section{Recent developments}
\label{sec:recent}
While ongoing efforts aim to further improve \mcmule{}’s QED predictions, the focus here is on extending \mcmule{} beyond pure QED.

\subsection{Bringing hadrons into the loop}
\label{sec:hadrons}
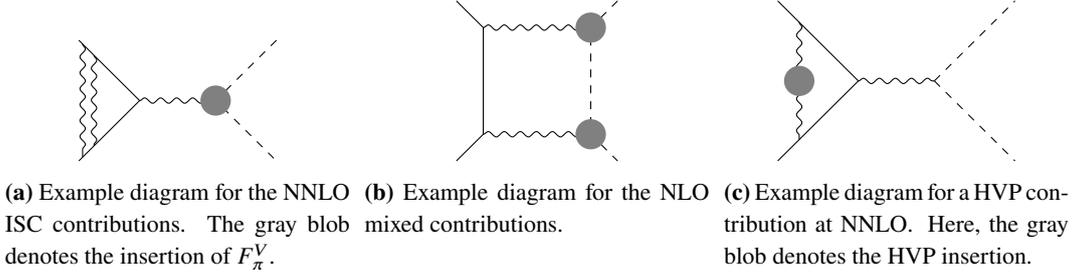
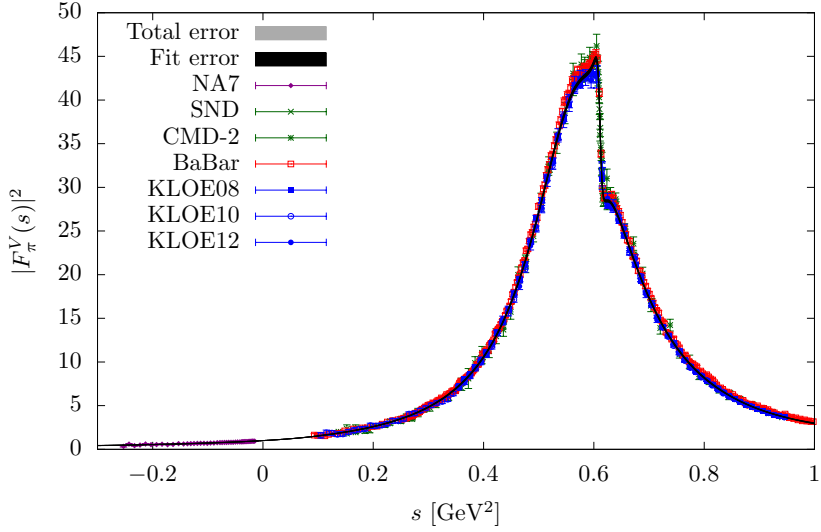
\begin{figure}
    \centering
    \begin{subfigure}[t]{0.3\textwidth}
        \centering
        \begin{tikzpicture}
            \draw (-1.3,0.8) -- (-0.5,0) -- (-1.3,-0.8);
            \draw[photon] (-0.5,0) -- (0.5,0);
            \draw[photon] (-1.1,0.6) -- (-1.1,-0.6);
            \draw[photon] (-1.25,0.75) -- (-1.25,-0.75);
            \draw[dashed] (1.3,0.8) -- (0.5,0) -- (1.3,-0.8);
            \fill[gray] (0.5,0.) circle (0.2);
        \end{tikzpicture}
        \caption{Example diagram for the \ac{NNLO} \acs{ISC} contributions.
        The gray blob denotes the insertion of $F^V_\pi$.}
        \label{fig:NNLOISC}
    \end{subfigure}%
    ~ 
    \begin{subfigure}[t]{0.3\textwidth}
        \centering
        \begin{tikzpicture}
            \draw (135:1.5) -- (135:1) -- (-135:1) -- (-135:1.5);
            \draw[dashed] (45:1.5) -- (45:1) -- (-45:1) -- (-45:1.5);
            \draw[photon] (135:1) -- (45:1);
            \draw[photon] (-135:1) -- (-45:1);
            \fill[gray] (45:1) circle (0.2);
            \fill[gray] (-45:1) circle (0.2);
        \end{tikzpicture}
        \caption{Example diagram for the \ac{NLO}  mixed  contributions.}
        \label{fig:pionbox}
    \end{subfigure}%
    ~ 
    \begin{subfigure}[t]{0.3\textwidth}
        \centering
            \begin{tikzpicture}
            \draw (135:1.5) -- (0,0) -- (-135:1.5);
            \draw [photon] (135:1.1) -- (-135:1.1);
            \fill[gray] ({cos(135)*1.1}, 0) circle (0.2);
            \draw [photon] (0,0) -- (1,0);
            \draw [dashed] (1,0) --+ (45:1.5);
            \draw [dashed] (1,0) --+ (-45:1.5);
        \end{tikzpicture}
        \caption{Example diagram for a \ac{HVP} contribution at \ac{NNLO}. Here, the gray blob denotes the \ac{HVP} insertion.}
        \label{fig:HVPnnlo}
    \end{subfigure}
    ~ 
    \begin{subfigure}[t]{0.75\textwidth}
        \centering
        \includegraphics[width=\textwidth]{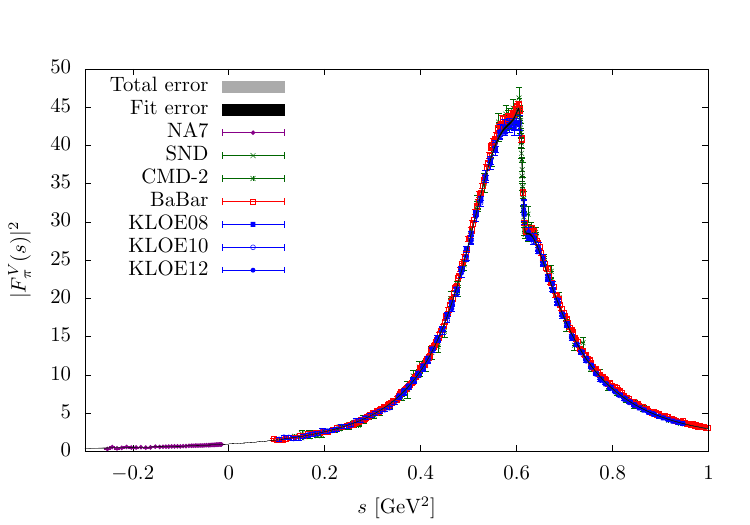}
        \caption{Fit of (absolute square of) the pion \ac{VFF} based on several measurements. Figure taken from~\cite{Colangelo:2018mtw}.}
        \label{fig:pionVFF}
    \end{subfigure}%
    \caption{Example contributions and hadronic data input for $ee\to\pi\pi$.}
    \label{fig:hadronicthings}
\end{figure}

When performing precision calculations at centre-of-mass energies up to a few GeV, particular care needs to be devoted to the treatment of non-perturbative input.
This includes the \ac{HVP}, form factors like the \ac{VFF} of the pion $F^V_\pi(q^2)$, or even more complicated objects needed for example for hadronic light-by-light contributions.
The pion \ac{VFF} and its non-trivial dependence on $q^2$ is shown in Figure~\ref{fig:pionVFF}.
As long as such hadronic objects only enter tree-level amplitudes, for example for the process $ee\to\pi\pi$ at \ac{LO} and for all its \ac{ISC} (shown in the diagram in Figure~\ref{fig:NNLOISC}), they can be included by a simple multiplication.
However, once they appear in loops, such as in the diagram in Figure~\ref{fig:pionbox}, their momentum dependence enters the loop integrals, which can no longer be evaluated using standard techniques.
Bringing the resulting integral into a standard form is possible by employing a (once-subtracted) dispersion relation, here given for $F^V_\pi(q^2)$,
\begin{align}
    \frac{F^V_\pi(q^2)}{q^2} = \frac{F^V_\pi(0)}{q^2} - \frac{1}{\pi} \int^\infty_{4 m^2_\pi} \frac{\mathrm{d} s_1}{s_1} \frac{\Im \, F^V_\pi(s_1)}{q^2 - s_1+ i \delta} \quad \text{with} \quad F^V_\pi(0) = 1\,.
    \label{eq:disprelation}
\end{align}
Once~\eqref{eq:disprelation} is applied, the loop integral can be evaluated analytically with standard methods, while the integral over the dispersive parameter $s_1$ starting at the threshold $s_{\rm thr} = 4 m^2_\pi$ is performed numerically~\cite{Cabibbo:1961sz}.
The drawbacks of~\eqref{eq:disprelation} are apparent:
\begin{itemize}
    \setlength\itemsep{0em}
    \item The integral has to be performed until $s_1 \to \infty$. 
    This challenges the numerical stability of the integrand.
    \item The dispersion relation~\eqref{eq:disprelation} changes the propagator structure of the resulting terms.
    The photon propagator is replaced by a propagator with mass $\sqrt{s_1}$, a \emph{disperon}, rendering the calculation of the required matrix elements tedious. 
\end{itemize}
In order to overcome those, we rely on \emph{disperon QED}~\cite{Fang:2025mhn}, a method to deal with hadronic input in loop processes in Monte Carlo codes.
The method is based on three pillars:
\begin{itemize}
    \setlength\itemsep{0em}
    \item[i)] The use of \OpenLoops{}:
    The calculation of the resulting amplitudes is delegated to \OpenLoops{}, for which a dedicated disperon QED model containing disperons, their interaction vertices and renormalisation has been implemented.
    \item[ii)] Disperon effective theory: In the regime of very large values of $s_1$, numerical stability and computational efficiency are enhanced by adopting an effective description motivated by \ac{EFT} arguments. 
    This purpose-built \ac{EFT}, disperon effective theory, is constructed by integrating out the heavy disperon and retaining contributions up to a fixed order in the inverse mass expansion.
    Schematically, the integral of~\eqref{eq:disprelation} gets divided into two regions whose matrix elements are calculated using different methods
    \begin{align}
        \int^{s_{\rm cut}}_{4 m^2_\pi} \frac{\mathrm{d} s_1}{s_1} \,\text{\OpenLoops{}} +  \int^\infty_{s_{\rm cut}} \frac{\mathrm{d} s_1}{s_1}\,\text{disperon effective theory}\,.
    \end{align}
    The cut off parameter $s_{\rm cut}$ is chosen such that the final result is independent of it within the numerical error of integration.
    This has to be verified explicitly for every kinematic scenario.
    An example for the observable 
    \begin{align}
        \label{eq:defthavg}
        \thavg = \frac{(\theta^- - \theta^+ + \pi)}{2}\,,
    \end{align}
    where $\theta^\pm$ is the angle of the $\pi^\pm$ is shown in Figure~\ref{fig:asym}.
    Here, a kinematical configuration close to the CMD-3 experiment at $\sqrt{s} = 0.7\,{\rm GeV}$ with
    \begin{align}
    \begin{split}
        |\vec p_\pm| &> 0.45\times\sqrt{s}/2\,,\\
        \delta\phi &= \big||\phi^+-\phi^-|-\pi\big| < 0.15\,{\rm rad}\,,\\
        \xi &= \big| \theta^++\theta^--\pi\big| < 0.25\,{\rm rad}\,,
    \end{split}
\end{align}
where $\phi^\pm$ is the azimuthal angle and $\vec p_\pm$ denote the three-momenta of the $\pi^\pm$, is used.
Figure~\ref{fig:asym} shows the differential distribution for $\thavg$ of $ee\to \pi\pi$ at \ac{LO} as well as available contributions at \ac{NLO} and \ac{NNLO}.
At \ac{NLO}, the mixed corrections with \ac{VFF} insertions into the loop integral (see Figure~\ref{fig:pionbox}), depend on the value of $s_{\rm cut}$.
Figure~\ref{fig:asym} shows these contributions with different values.
While the result clearly changes for $s_{\rm cut} \lessapprox 0.5\,{\rm GeV}^2$, it stabilises at higher $s_{\rm cut}$.
In this case, we have chosen $s_{\rm cut} = 4\,{\rm GeV}^2$ for the final calculation.
\begin{figure}
    \centering
    \includegraphics[width=0.8\textwidth]{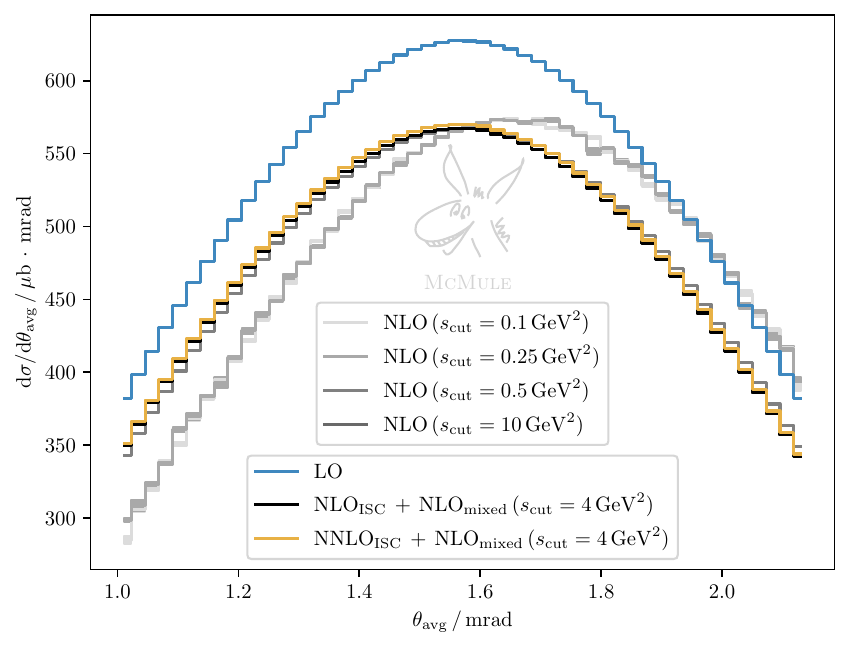}
    \caption{
    Differential distribution for $\thavg$ as defined in~\eqref{eq:defthavg}.
    The mixed \ac{NLO} contributions that are computed using disperon QED are shown for different values of $s_{\rm cut}$.
    The physics run~\cite{Fang:2025mhn} is performed using $s_{\rm cut} = 4\,{\rm GeV}^2$ and indicated using solid colours.
    The partial \ac{NLO} and \ac{NNLO} results include the full \ac{LO}.
    The pion \ac{VFF} was taken from~\cite{Colangelo:2018mtw} and is shown in Figure~\ref{fig:pionVFF}.}
    \label{fig:asym}
\end{figure}
    \item[iii)] Threshold subtraction: In $s$-channel processes, a singularity arises in the dispersive integral when $s_1$ equals the momentum transfer $q^2$.
    This can be overcome by formulating a general description of the singular behaviour and its associated counterterm, subtracting it from the amplitude prior to numerical integration and restoring it in analytically integrated form.
    We find that the singularity at $q^2 = s_1$ is given by a universal description
    \begin{align}
        \frac{1}{s_1} \left( \frac{q^2}{s_1 - q^2 - i \delta} \right)^{1+2\epsilon} f(\text{kinematics})\,,
    \end{align}
    where the universal function $f$ only depends on the kinematics of the process.
\end{itemize}
By applying these methods, \mcmule{} successfully performed a \ac{NLO} calculation of $ee\to\pi\pi$~\cite{Fang:2025mhn} where the pion-photon interactions are modelled in \ac{FsQED}\footnote{We refer to~\cite{Fang:2025mhn,Aliberti:2024fpq} for a discussion about the applicability of and further information about \ac{FsQED}.}~\cite{Colangelo:2014dfa,Colangelo:2015ama,Colangelo:2022lzg}, i.e. in \acl{sQED} with a pion \ac{VFF} at every interaction as shown in Figure~\ref{fig:pionbox}.
An example observable, $\thavg$, obtained using disperon QED is shown in Figure~\ref{fig:asym}.
The mixed \ac{NLO} corrections to this quantity have been studied extensively in~\cite{Budassi:2024whw,Ignatov:2022iou}, because their asymmetric structure makes them a particularly sensitive test of the treatment of radiative corrections in $ee\to\pi\pi$.

We emphasise that the discussion above for treating $F^V_\pi(q^2)$ at \ac{NLO} can be applied in the same way as dealing with \ac{HVP} insertions at \ac{NNLO} (see Figure~\ref{fig:HVPnnlo}).
In that case, the dispersion relation~\eqref{eq:disprelation} gets modified to, e.g.~\cite{Jegerlehner:2017gek}
\begin{align}
    \frac{\Pi_{\gamma\gamma}(q^2)}{q^2} = - \frac{1}{\pi} \int^\infty_{4 m^2_\pi} \frac{\mathrm{d} s_1}{s_1} \frac{\Im\,\Pi_{\gamma\gamma}(s_1)}{q^2 - s_1 + i \delta} \quad \text{with} \quad \Pi_{\gamma\gamma}(0) = 0 \,,
\end{align}
where $\Pi_{\gamma\gamma}(q^2)$ denotes the current-current correlator of either the hadronic part of the self-energy
\begin{align}
    \label{eq:sigmaPi}
    \Sigma_{XX}(q^2) = \Sigma_{XX}(0) + q^2 \, \Pi_{XX}(q^2) \quad \text{with} \quad X = \gamma\,,
\end{align}
obtained from $e^+ e^- \to \text{hadrons}$ measurements or its resummation.
We further emphasise that the methods are not restricted to the \ac{HVP}, but can equally be applied to the perturbative vacuum polarisation, effectively reducing a two-loop integral to a one loop one.

\subsection{Adding electroweak effects}
\label{sec:EW}
\begin{figure}
    \centering
    \includegraphics[width=0.8\textwidth]{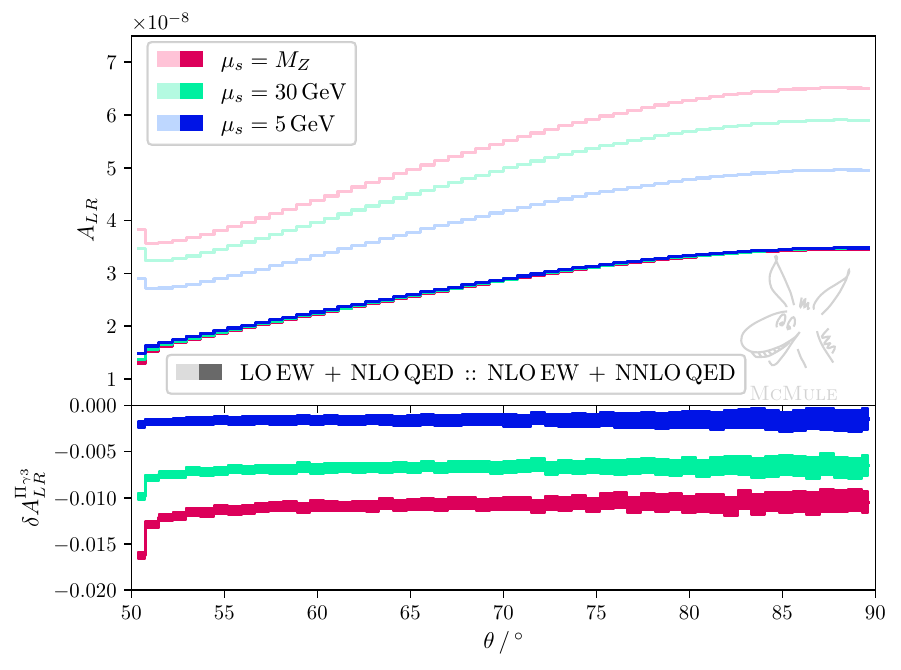}
    \caption{$A_{LR}$ for the MOLLER experiment as a function of $\theta$ defined in~\eqref{eq:CoMtheta} for three choices of $\mu_s$ in \ac{RGE}-improved perturbation theory.
    The bottom panel shows the impact of using a different model for the hadronic $\Pi_{\gamma 3}$ as defined in~\eqref{eq:HVPimpact}.}
    \label{fig:ALRmoller}
\end{figure}

While \ac{EW} effects are highly suppressed at lower energies, they are the key for parity-violating observables, such as the asymmetry between right- and left-handed electrons scattering off different targets
\begin{align}
    A_{LR} = \frac{\sigma_L - \sigma_R}{\sigma_L + \sigma_R}\,.
\end{align}
The upcoming MOLLER experiment~\cite{MOLLER:2014iki} aims to measure $A_{LR}$ in M{\o}ller scattering $e^-e^-\to e^- e^-$ at $\sqrt{s} = 106 \,{\rm MeV}$, an observable highly sensitive to the weak mixing angle $s_W$ and to radiative corrections, whose precise control is therefore essential.
While the \ac{NLO} corrections can easily reach $40\%$, full \ac{NNLO} corrections including resummation of large logarithms of the form $\log(\mu_{s}/\mu_{h})$, where $\mu_s \sim \{s, m^2_e, \dots\} $ and  $\mu_h \sim \{M^2_Z, M^2_W, \dots\}$, are still missing. 
\mcmule{}’s recent effort~\cite{Kollatzsch:2025pnp} introduces a framework designed to meet this precision challenge.
It combines \ac{NNLO} QED with \ac{NLO} \ac{EW} effects in \ac{LEFT}~\cite{Jenkins:2017jig,Dekens:2019ept} up to and including mass dimension 6 to fully exploit the scale hierarchy $\mu_s \ll \mu_h$.
A calculation in \ac{LEFT} has the following advantages:
\begin{itemize}
    \setlength\itemsep{0em}
    \item The integrals appearing in \ac{LEFT} are substantially simpler than in the full \ac{SM}. 
    Although this simplification is of little relevance at one loop, it becomes crucial at two loops.
    \item \ac{LEFT} offers a natural framework to consistently resum all large logarithms of the form $\log(\mu_{s}/\mu_{h})$. 
    Given the ongoing efforts, e.g.~\cite{Aebischer:2025hsx,Naterop:2025cwg}, a resummation at \ac{NLL} accuracy appears feasible.
    \item All \ac{EW} effects are encoded in the Wilson coefficients of \ac{LEFT} and the matrix elements in \mcmule{} are implemented as a function of them.
    This allows for a flexible use to study for example different input and renormalisation schemes, \ac{RGE} effects as well as new physics scenarios -- all in one implementation.
\end{itemize}
The upper panel of Figure~\ref{fig:ALRmoller} shows $A_{LR}$ for the kinematical situation of the MOLLER experiment with longitudinal polarisation $P = 90\%$ and $50^\circ \le \theta_{3,4} \le 130^\circ$ (centre-of-mass-frame)
as a function of the scattering angle
\begin{align} \label{eq:CoMtheta}
    \theta &= \frac{1}{2} \left(\pi - \vert \theta_3 - \theta_4 \vert\right) \,, 
\end{align}
that takes into account that the two scattered electrons described by $\theta_3$ and $\theta_4$ are indistinguishable. 
On top of \ac{NNLO} QED and \ac{NLO} \ac{EW} effects, Figure~\ref{fig:ALRmoller} includes the \acs{LL} running of the Wilson coefficients up to the scale $\mu_s$.
As evident from the figure, the resummation of these logarithms is crucial. 
While the dependence on $\mu_s$ is strongly reduced at \ac{NLO}, the combination of \ac{NNLO} \ac{EW} with \ac{NLL} running is required for a fully satisfactory description.

Another challenge when dealing with \ac{EW} effects at low energies is the evaluation of non-perturbative contributions to the $\gamma - Z$ mixing.
Expressed in terms of current-current correlators (see also~\eqref{eq:sigmaPi}), this reads 
\begin{align}
\label{eq:gammaZpi}
    \Pi_{\gamma Z}(q^2) = \frac{1}{c_W s_W} \left(\Pi_{\gamma 3}(q^2) - s_W^2 \Pi_{\gamma \gamma}(q^2) \right)\,,
\end{align}
where
$\Pi_{\gamma 3}$ is related to the third component of the isospin current of the $Z$ coupling.
The combination $\Pi_{\gamma\gamma}(q^2) - \Pi_{\gamma\gamma}(0)$ enters standard QED calculations (see Section~\ref{sec:hadrons}, in particular Figure~\ref{fig:HVPnnlo}) and is straightforwardly extracted from $e^+e^- \to \text{hadrons}$ data. 
It is therefore available for Monte Carlo applications through several dedicated implementations~\cite{Keshavarzi:2019abf,Ignatov:hvp,Jegerlehner:hvp19}.
In contrast, the evaluation of the hadronic $\gamma$–$Z$ mixing remains a recurring subject in the literature and may limit the precision of the affected observables.
At the moment, only {\tt alphaQED}~\cite{Jegerlehner:hvp19} is able to provide predictions for the quantity $\Pi_{\gamma 3}(q^2)-\Pi_{\gamma 3}(0)$ that contributes to $\Pi_{\gamma Z}$.
The main idea here is to perform a flavour separation of $e^+ e^- \to \text{hadrons}$ data and then recombine the individual re-weighted contributions using the Vector Meson Dominance model~\cite{Jegerlehner:2017zsb}.
In the second panel of Figure~\ref{fig:ALRmoller}, we show the impact on $A_{LR}$ of using a model for $\Pi_{\gamma 3}$ different from {\tt alphaQED}.
To that end, we define the quantity
\begin{align}
    \label{eq:HVPimpact}
    \delta A_{LR}^{\Pi_{\gamma 3}}
    \equiv 
    \frac{
    A_{LR}(\Pi_{\gamma 3}\,{\rm via\,\texttt{alphaQED}}) 
    - A_{LR}(\Pi_{\gamma 3}\,{\rm via\,\cite{Hoferichter:2025yih}})
    }{
    A_{LR}(\Pi_{\gamma 3}\,{\rm via\,\cite{Hoferichter:2025yih}})
    }\, .
\end{align}
This model~\cite{Hoferichter:2025yih,Hoferichter:privateHVPcode}, valid only in the space-like region, relies on {\tt alphaQED} only for $\Pi_{\gamma\gamma}$ and uses recent lattice input to describe $\Pi_{\gamma 3}$.
While the impact described by~\eqref{eq:HVPimpact} reaches values between $1 - 1.5\%$ at $\mu_s = M_Z$ and thus substantially contributes to the theoretical uncertainty, it reduces to subpercent level for the preferred, smaller values of $\mu_s$. 
Given the targeted experimental precision of the MOLLER experiment of about $2\%$~\cite{MOLLER:2014iki}, the specific choice of the $\Pi_{\gamma 3}$ model, when used in an \ac{RGE}-improved setting, has no impact on the prediction of $A_{LR}$.

\section{Plans for the future}
\label{sec:conclusions}
The recent effort~\cite{Kollatzsch:2025pnp}, described in Section~\ref{sec:EW}, lays the groundwork for a systematic inclusion of \ac{EW} effects at $\sqrt{s} \ll M_Z$ within a Monte Carlo framework at the targeted experimental precision.
Building on this foundation, the extension to \ac{NNLO} \ac{EW} accuracy, including a two-loop calculation in \ac{LEFT} as well as the matching of the \ac{LEFT} to the \ac{SM} at two loops, is currently in progress.
As discussed in these proceedings,
the uncertainty due to different $\Pi_{\gamma 3}$ models is small, so pushing for \ac{NNLO} remains a well-motivated next step.

While the methods described in Section~\ref{sec:hadrons} were illustrated for the process $ee\to\pi\pi$, they were developed with more complicated processes in mind. 
Applying them to $ee\to\pi\pi\gamma$ is ongoing work.
The \mcmule{} effort to improve predictions for $ee\to\pi\pi(\gamma)$ (including $ee\to ee(\gamma)$ and $ee\to\mu\mu(\gamma)$) is part of a bigger community effort: RadioMonteCarLow2~\cite{Aliberti:2024fpq,RMCLW2:website}, whose aim it is to maintain, preserve accessibility to, and further develop existing Monte Carlo codes relevant for $e^+ e^- \to \text{hadrons}$ at a few GeV.
The \ac{HVP} contributions to $ee\to\mu\mu\gamma$ at \ac{NNLO} as well as their resummation will be computed using disperon QED.

Beyond pions, disperon QED can be employed to describe proton form factors in lepton-proton scattering.
When combined with the \ac{EW} framework described in Section~\ref{sec:EW}, \mcmule{} has the potential to provide theoretical support for the P2 experiment~\cite{Becker:2018ggl} which aims to measure $A_{LR}$ in $e p \to e p$.

\acknowledgments{I acknowledge support from the Swiss National Science Foundation (SNSF) under grant \href{https://data.snf.ch/grants/grant/207386}{207386}.
A huge thank you to all collaborators of~\cite{Kollatzsch:2025pnp,Fang:2025mhn}, to Martin Hoferichter for insightful discussions and for sharing the $\Pi_{\gamma Z}$ code~\cite{Hoferichter:privateHVPcode}, to Yizhou Fang for performing the runs with different $s_{\rm cut}$ values and preparing a first version of Figure~\ref{fig:asym}, and to the entire \href{https://mule-tools.gitlab.io/team.html}{\mcmule{}
 team} for making this framework possible.
}
\bibliographystyle{JHEP}
\bibliography{references}

\end{document}